\documentclass[aps,prd]{revtex4}
\usepackage{graphicx}
\newcommand{\intq}{\int\frac{d^3q}{(2\pi)^3}}

\newcommand{\no}{\nonumber}

\newcommand{\wt}{\widetilde}

\def\a{\alpha}
\def\b{\beta}
\def\d{\delta}
\def\e{\epsilon}
\def\p{\phi}

\def\l{\lambda}
\def\m{\mu}
\def\n{\nu}
\def\r{\rho}

\def\th{\theta}
\def\o{\omega}

\def\G{\Gamma}

\def\be{\begin{equation}}
\def\ee{\end{equation}}
\def\bea{\begin{eqnarray}}
\def\eea{\end{eqnarray}}
\def\bi{\begin{itemize}}
\def\ei{\end{itemize}}
\def\bc{\begin{center}}
\def\ec{\end{center}}
\def\no{\nonumber}
\def\cp{{\cal P}}
\def\ck{{\cal K}}
\begin{document}
\title{Induced magnetic moment in noncommutative
Chern-Simons scalar QED}

\author{Prasanta K. Panigrahi\footnote{prasanta@prl.ernet.in}$^a$ and
T. Shreecharan\footnote{shreet@prl.ernet.in}$^{a,b}$}

\affiliation{$^a$ Physical Research Laboratory, Navrangpura,
Ahmedabad-380 009, India\\
$^b$ School of Physics, University of Hyderabad, Hyderabad-500
046, India}

\begin{abstract}
We compute the one loop, $O(\th)$ correction to the vertex in the
noncommutative Chern-Simons theory with scalar fields in the
fundamental representation. Emphasis is placed on the parity odd
part of the vertex, since the same leads to the magnetic moment
structure. We find that, apart from the commutative term, a
$\th$-dependent magnetic moment type structure is induced. In
addition to the usual commutative graph, cubic photon vertices
also give a finite $\th$ dependent contribution. Furthermore, the
two two-photon vertex diagrams, that give zero in the commutative
case yield finite $\th$ dependent terms to the vertex function.
\end{abstract}

\maketitle

\section{Introduction}

The possibility of particles carrying fractional angular momentum
on a plane is, by now, well accepted \cite{wilczek}. The role of
Chern-Simons (CS) term \cite{schonfeld} in inducing fractional
spin has been carefully investigated \cite{tze}. Field
theoretically, it has been shown in 2+1 dimensions that, one can
calculate fractional angular momentum eigenvalues of single
particle states. Furthermore, Polyakov showed that, the
interaction of scalar particles with the CS gauge field leads to
the transmutation of a boson into a spinning particle
\cite{polyakov}. An interesting consequence of this is the
appearance of spin magnetic moment for the bosons, not possible in
3+1 dimensions. Although not present at the tree level, the boson
spin is induced at the one-loop level, leading to a magnetic
moment for the bosons \cite{kogan2}. The existence of magnetic
moment leads to unusual planar dynamics, as shown for scalars and
spinors in the context of Maxwell-Chern-Simons electrodynamics
\cite{kogan1,wallet}. Therefore, the magnetic moment of anyons has
been studied extensively \cite{rashmi}. CS field theories have
attracted considerable attention in the literature \cite{dunne} as
effective theories for explaining the physics of fractional Hall
effect \cite{stone}.

Noncommutative (NC) field theories have been generating interest
in the past few years in the context of string theory
\cite{witten}. Though the idea of noncommutativity, as a regulator
for the divergences in field theory, was introduced very early on
\cite{snyder}; only recently, has it been been taken up as an
independent field of study. NC field theories are defined on a
manifold with coordinates that do not commute:
$[x^\m,x^\n]=i\th^{\m\n}$. These theories have attracted
considerable attention in the context of quantum Hall effect
\cite{bellissard}. The NCCS theory and its variants have been
quite useful in explaining the filling fraction of the electrons
in the lowest Landau level \cite{susskind}; there are a number of
other physical situations where noncommutative field theories have
been useful \cite{nekrasov}.

It has been shown that even though the CS term is absent at the
tree level, it is generated due to radiative corrections at the
one loop level in the presence of fermions
\cite{redlich,chu,chandra}. A CS type term is also generated in
the effective action of charged particles in a magnetic field
\cite{sakita}. Recently, various aspects of NC theories with a CS
term have been under the scrutiny of a number of authors
\cite{bichl,adas1,ncmartin,chen,grandi}.

Keeping this as well as the fact that, a spin magnetic moment can
play an important role in the planar dynamics, we compute the
magnetic moment of scalar particles in the context of
noncommutative scalar QED in 2+1 dimension with a tree level CS
term. It can be noticed from the action that, apart from the usual
vertex a three gluon vertex also contributes even for an $U(1)$
theory. This feature is quite similar to the non-Abelian
commutative theories \cite{nair}. The two-photon Feynman diagram
will also contribute due to the appearance of non-planar
integrals. These NC contributions to the vertex vanish when the NC
parameter $\th$ is set to zero giving the commutative result.

The paper is organized as follows. In the following section, we
introduce the NCCS action with bosonic matter fields and state the
Feynman rules stemming from it. In Section III, the vertex
contributions arising from all the diagrams \cite{binosi} at
one-loop level are computed. We concentrate on the parity odd
gauge invariant pieces, since the same lead to magnetic moment
type interactions. Up to first order in $\th$,  the  vertex
expression is found to contain real and imaginary pieces. The
imaginary part depends on the magnitude of the non-commutative
parameter. Contributions indicating $\th$ dependent spin type
structures are identified, akin to 3+1 dimensional theories.
Conclusions are presented in section IV, pointing out future
directions of work. For the sake of completeness, we provide in
the appendix the results for non-planar momentum integrals,
encountered in the course of our calculation.

\section{Feynman rules}

\begin{figure}
\begin{tabular}{cc}
\begin{minipage}{3in}
\includegraphics[width=1.5in,height=1in]{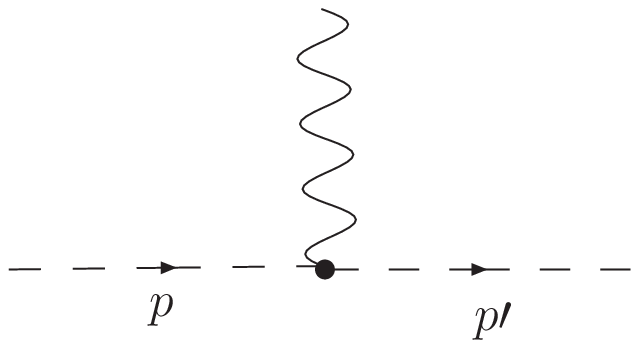}
\caption{The boson-photon vertex.} \label{sf1}
\end{minipage}
& {\large $\equiv ie(p+p^{\prime})_\m \exp\left[\frac{i}{2}p
\times p^{\prime} \right]$.} \\
\begin{minipage}{3in}
\includegraphics[width=1.5in,height=1in]{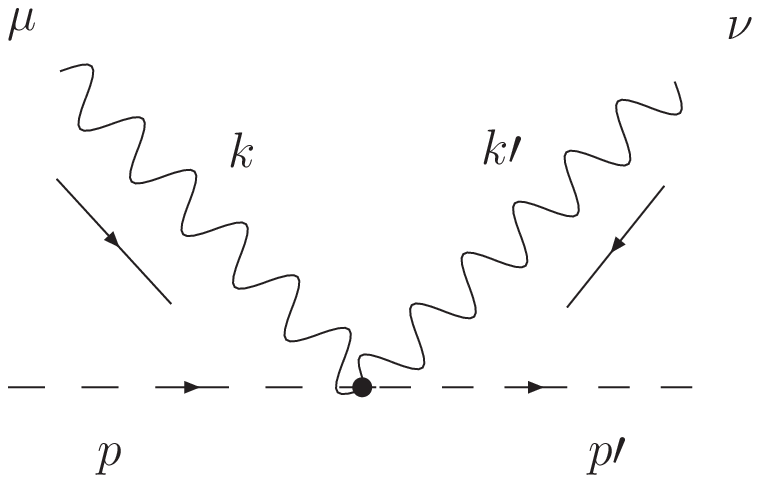}
\caption{The two scalar-photon vertex.} \label{sf2}
\end{minipage}
& {\large $\equiv 2ie^2 g^{\m\n} \exp\left[\frac{i}{2}p \times
p^{\prime} \right] \cos\left[(k \times k^{\prime})/2\right]$.} \\
\begin{minipage}{3in}
\includegraphics[width=1.5in,height=1in]{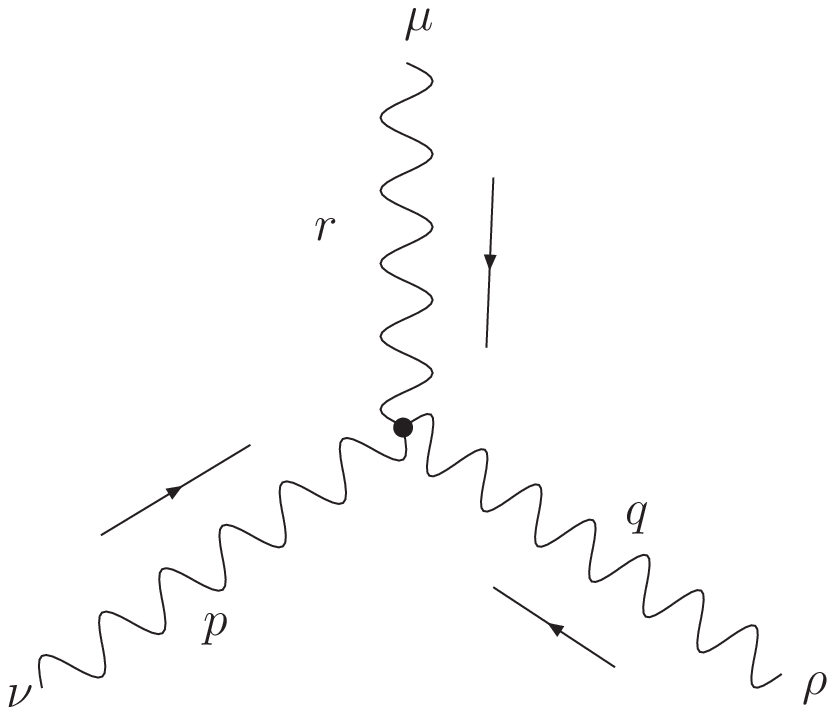}
\caption{The three gauge boson vertex.} \label{sf3}
\end{minipage}
& {\large $\equiv 2ieM\e^{\n\m\r} \sin\left[(p \times
r)/2\right]$.}
\end{tabular}
\end{figure}
The $U(1)$ NCCS action with scalar matter fields is given by
    \be \label{action}
S_{\star}= \int d^3x \left[\frac{M}{2}\e^{\m\n\r}
\left(A_{\m}\star\partial_{\n}A_{\r}+ \frac{2ie}{3} A_{\m} \star
A_{\n} \star A_{\r} \right) + ({D_\m\p}^{\dagger}\star D^\m \p -
m\p^{\dagger}\star\p) \right],
    \ee
where $D_\m \phi=\partial_\m\phi - ieA_\m \star\phi$. The matter
field has been taken to be in the fundamental representation. The
presence of $(A_\m \star A_\n \star A_\r)$ term leads to
self-interaction amongst the photons and is similar to the
commutative non-Abelian version of the theory. Furthermore,
two-photon terms from the  matter action also contribute to the
vertex, which give zero in the commutative version of the above
theory. As is already known \cite{nekrasov}, the propagators
retain their structure in the NC theories and only the vertices
are modified. Therefore, the gauge field and the scalar
propagators for the above NC action are
    \be \label{propagator}
G_{\m\n}(p)= - \frac{1}{M}\e_{\m\n\r}\frac{p^{\r}}{p^2}\quad {\rm
and} \quad D(p)=\frac{i}{p^2-m^2+i\e},
    \ee
respectively. Here the gauge propagator is defined in the Landau
gauge, since it is known to be infrared safe \cite{pisarski}. The
interaction vertices and their expressions are shown in Figs.
(\ref{sf1}), (\ref{sf2}), and (\ref{sf3}). In the expressions for
the figures and in what follows, $()\times ()\equiv\th^{\m\n}()_\m
()_\n$.

\section{Induced magnetic moment}

In this section, we evaluate various scalar one-loop diagrams
contributing to the vertex, upto first order in $\th$. The
calculations have been broken up into different subsections,
corresponding to different diagrams, for the sake of convenience.

\subsection{Boson-photon vertex contribution}

\begin{figure}
\begin{center}
\includegraphics[width=2.5in,height=2in]{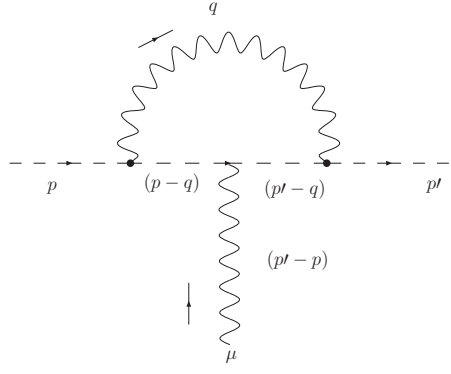}
\end{center}
\caption{scalar-gauge field vertex.} \label{sv1}
\end{figure}

The contribution to the vertex arising from the diagram shown in
Fig. (\ref{sv1}), which is also present in the commutative case,
can be written in the form
    \be \label{gscalar}
\G^1_\m = -e^2 \int \frac{d^3q}{(2\pi)^3}{(p+p^{\prime}-2q)_\m
(2p-q)_\n (2p^\prime -q)_\r G^{\n\r}(q) \over
[(p-q)^2-m^2][(p^\prime -q)^2 -m^2]}\, e^{-iq \times
\ck}e^{\frac{i}{2}p\times p^{\prime}},
    \ee
where $\ck_\m\equiv(p^{\prime}-p)_\m$. The above can be simplified
to yield
    \be
\G^1_\m= -\frac{4e^2}{M}\intq {\e^{\n\r\a}p_\n {p^{\prime}}_\r
q_\a(p+p^{\prime}-2q)_\m \over q^2[(p-q)^2-m^2][(p^\prime -q)^2
-m^2]}\, e^{-iq \times \ck}e^{\frac{i}{2}p\times p^{\prime}}.
    \ee
The loop integral can be evaluated in the standard manner. After
combining the denominators [we have used Eq. (\ref{int2})] and
shifting the integration variable we get
    \be
\G^1_\m = -\frac{8e^2}{M} \int_0^1 dx \int_0^x dy\intq
{\e^{\n\r\a}p_\n p^{\prime}_\r {\tilde
q}_\a[(2x-1)p_\m+(1-2x+2y)p^{\prime}_\m-2{\tilde
q}_\m]\over[{\tilde q}^2-\o_1^2]^3}\, e^{-i{\tilde q} \times
\ck}e^{-\frac{i}{2}(1-2y)p\times p^{\prime}},
    \ee
where $\o_1^2= (1-y)^2 m^2-(1-x)(x-y)\ck^2$ and $q={\bar
q}+(x-y)p^{\prime}+(1-x)p$. For the sake of notational simplicity,
we continue to denote the new integration variable ${\tilde q}$ as
$q$ in this, as well as later calculations. In solving the above
integrals, we retain only the $q_\m q_\a$ term, since only this
term gives magnetic moment type interaction. The momentum
integral, using the result of Eq. (\ref{int8}), yields
    \bea
\G^1_\m = -\frac{8ie^2\e_{\m\n\r}p^\n
{p^{\prime}}^\r}{(2\sqrt{\pi})^3M} \int_0^1 dx \int_0^x dy
e^{-\frac{i}{2}(1-2y)p\times p^{\prime}}
\left[\frac{|{\wt\ck}|}{2|\o_1|}\right]^{1/2}K_{1/2}(|{\wt\ck}|
|\o_1|).
    \eea
The parametric integrals can be handled in an elegant manner by
going to a particular frame of reference: the rest frame of the
scalar particle, where $p \times p^{\prime} = 0$. Also, we take
$\th^{0i}=0$, since it is known that space-time noncommutativity
violates unitarity \cite{mehen}. Using
    \be \label{bessel}
K_{\pm 1/2}(z) = \sqrt{\frac{\pi}{2}} \frac{e^{-z}}{\sqrt z},
    \ee
and retaining terms first order in $\th$ from the above expansion,
we get
    \be
\G^1_\m = -{ie^2\e_{\m\n\r}\cp^\n \ck^\r \over 4\pi
M}\left[\frac{1}{m}- \frac{|{\wt\ck}|}{2}\right].
    \ee
It must be mentioned that, the above expression is obtained in the
$\ck^2 \rightarrow 0$ limit. Furthermore, we have replaced $p$ and
$p^{\prime}$ using the relations for $\ck_\m$ and $\cp_\m=
{p^\prime}_\m+p_\m$.

\subsection{Three-photon vertex contribution}

\begin{figure}
\begin{center}
\includegraphics[width=2.5in,height=2in]{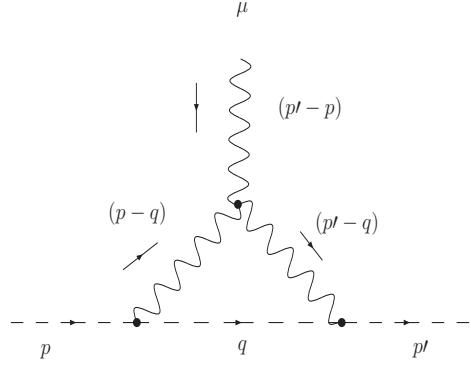}
\end{center}
\caption{The three photon vertex contribution.} \label{threep}
\end{figure}
Here, we deal with the three-gluon contribution shown in Fig.
(\ref{threep}), to the NC vertex:
    \bea \no
\G_\m^{2} &=& -2ie^2M \intq {(p^\prime+q)_\r (p-q)_\l
G^{\n\a}(p-q)\e_{\a\m\b}G^{\b\r}(p^\prime-q)\over(q^2-m^2)}\,
\sin\left[{(p-q)\times (p^\prime-p)\over 2}\right]\,
e^{\frac{i}{2}q\times \ck}, \\ \label{3g1}
    &=& \frac{2ie^2}{M}\intq
{\e^{\n\a\l}\e_{\m\a\b}\e^{\b\r\d} (p^\prime+q)_\r (p-q)_\l
(p+q)_\n (p^\prime-q)_\d \over (q^2 - m^2)(p-q)^2 (p^\prime -q)^2}
\, \sin\left[{p\times p^\prime - q\times \ck \over 2}\right]\,
e^{\frac{i}{2}q\times \ck}.
    \eea
The above vertex $\G_\m^{2}$, can be written in terms of planar
and non-planar contributions in the form,
    \be
\G_\m^{2} = \frac{4e^2}{M}\intq {\e^{\n\a\l}p_\l {p^\prime}_\a
q_\n q_\m \over (q^2 - m^2)(p-q)^2 (p^\prime -q)^2}
[e^{\frac{i}{2}p\times p^\prime}- e^{-\frac{i}{2}p\times p^\prime}
e^{iq\times \ck}].
    \ee
In obtaining the above expression, we have simplified the
numerator using the standard $\e$ manipulations. As before,
combining the denominators and shifting the integration variable
we get
    \be
\G_\m^{2}= \frac{8e^2}{M}\int^1_0 dx \int^x_0 dy \intq
{\e^{\n\a\l}p_\l {p^\prime}_\a q_\n q_\m \over
(q^2-\o_2^2)^3}\,[e^{\frac{i}{2}p\times p^\prime}- e^{iq\times
\ck} e^{\frac{i}{2}(1-2y) p\times p^\prime}],
    \ee
where $\o_2^2 = m^2y^2-(x-y)(1-x)\ck^2$. The vertex can be
separated as: $\G_\m^2=\G_\m^{2P}+\G_\m^{2NP}$. The planar part
can be simplified:
    \be
\G_\m^{2P}= -\frac{ie^2}{4\pi M}\int_0^1 dx \int_0^x dy
\frac{\e_{\m\a\l}{p^\prime}^\a p^\l}{my}.
    \ee
It can be noticed that the above planar contribution has a
logarithmic divergence. The non-planar contribution can be written
in the form
    \be
\G_\m^{2NP}= \frac{4ie^2\e_{\n\a\l}g^{\m\n}}{M(2\sqrt{\pi})^3}
\int_0^1 dx \int_0^x dy \left[\frac{|{\wt
\ck}|}{2|\o_2|}\right]^{1/2}K_{1/2}(|{\wt \ck}||\o_2|).
    \ee
On expanding the Bessel function and retaining contribution linear
in the NC parameter, we see that the log divergence from the
planar piece exactly cancels a similar divergence from the
non-planar contribution. Hence, the 3-photon vertex is divergence
free. Such a cancellation of divergences stemming from the planar
and non-planar contributions has been noted in the photon
self-energy calculation in 3+1 dimensions \cite{adas2}. The
contribution to the vertex can be combined into a compact form:
    \be
\G_\m^{2}= \frac{ie^2}{4\pi M} \e_{\m\n\r}{\cp}^\n {\ck}^\r
\frac{|{\wt \ck}|}{4}.
    \ee

\subsection{Two-photon vertices}

\begin{figure}
\begin{tabular}{cc}
\begin{minipage}{3in}
\includegraphics[width=2.5in,height=2in]{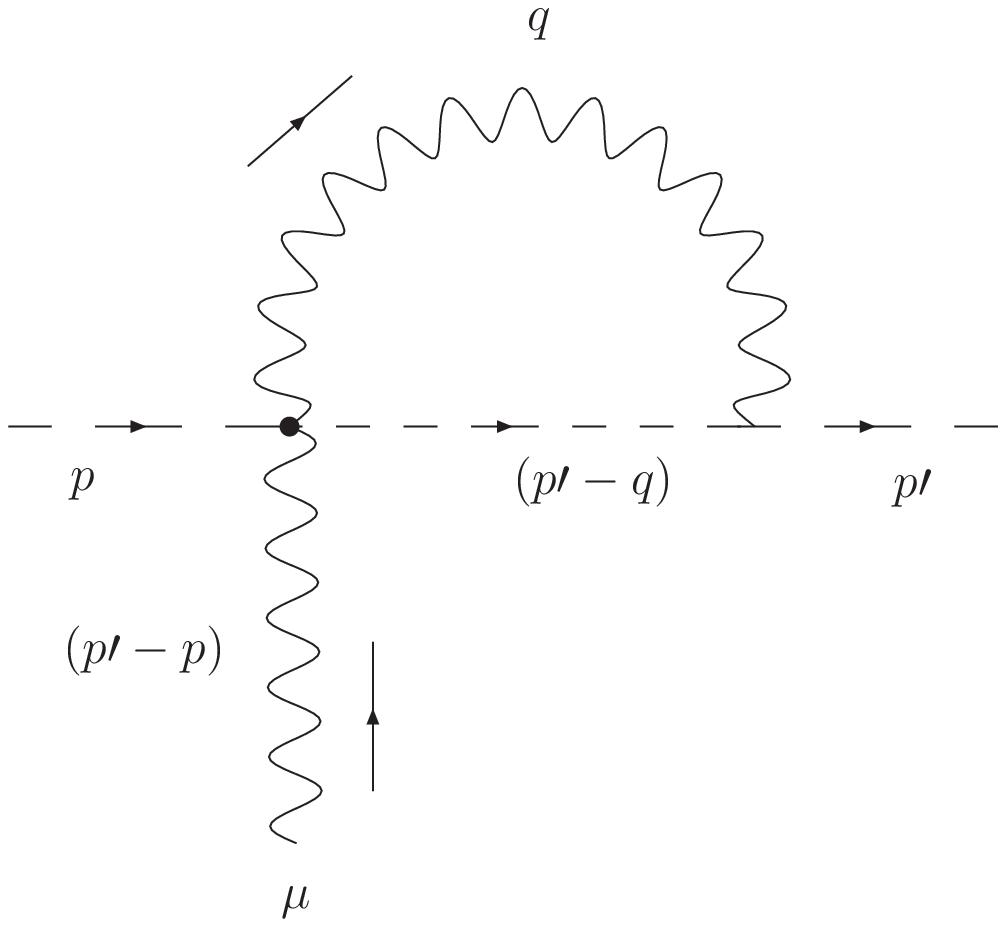}
\caption{The two photon vertex.} \label{twop1}
\end{minipage}
&
\begin{minipage}{3in}
\includegraphics[width=2.5in,height=2in]{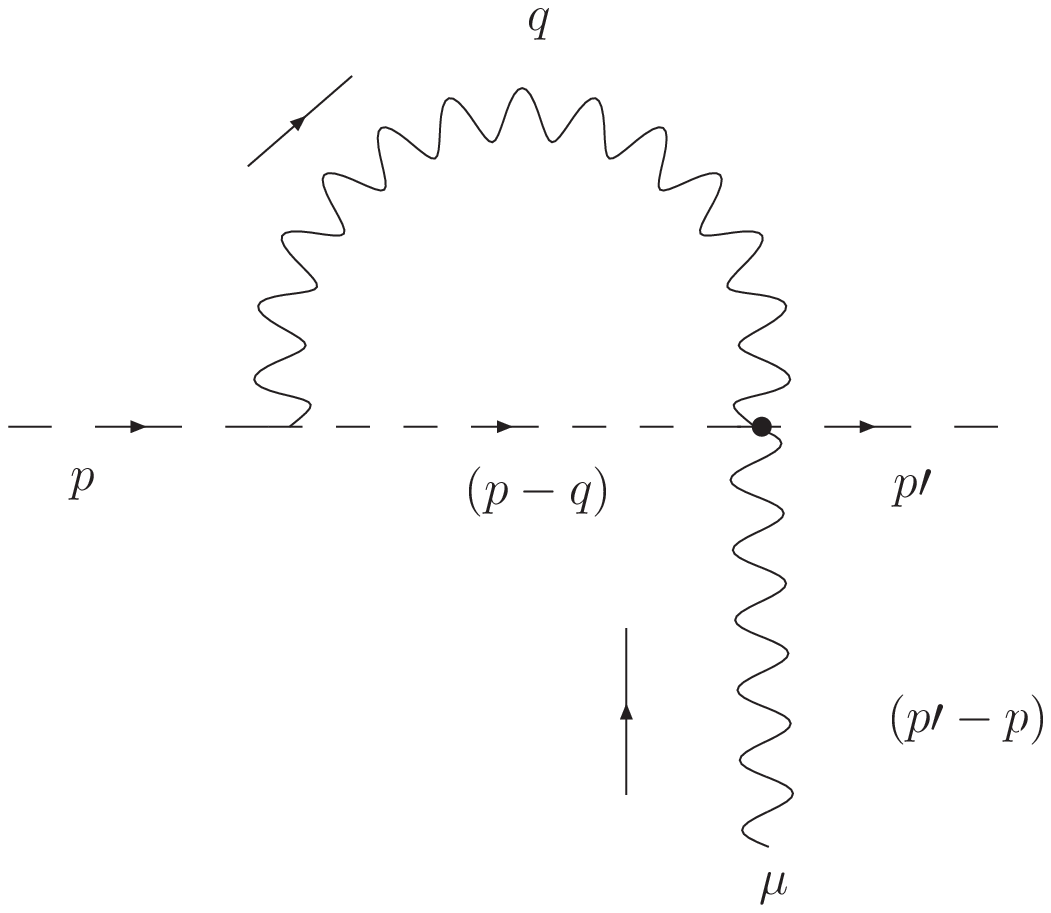}
\caption{The two photon vertex.} \label{twop2}
\end{minipage}
\end{tabular}
\end{figure}
The two photon vertex amplitude in Fig. (\ref{twop1}) can be
written in the form
    \be
\G_\m^{3} = \frac{2e^2}{M} \intq
{\e^{\n\r\l}g_{\m\n}(2p^{\prime}-q)_\r q_\l \over q^2[(p^{\prime}-
q)^2 - m^2]} \cos\left[\frac{q \times \ck}{2}\right]
e^{\frac{i}{2}p \times p^{\prime}} e^{-\frac{i}{2}q\times \ck},
    \ee
which yields,
    \be
\G_\m^{3}=\frac{2e^2}{M} \int_0^1 dx \intq
\frac{\e_{\m\r\l}{p^{\prime}}^\r q^\l}{(q^2
-\o_3^2)^2}\left[e^{\frac{i}{2}p \times p^{\prime}}+ e^{-i q\times
\ck} e^{-\frac{i}{2}(2x-1)p\times p^\prime}\right].
    \ee
In obtaining the above expression we have redefined the
integration variable by $q={\bar q}+ x p^\prime$ and defined
$\o_3^2= {p^\prime}^2 x^2$. It is clear that the planar
contribution is zero and only the non-planar integral survives:
    \be
\G_\m^{3NP}=\frac{2e^2\e_{\m\r\l}{p^\prime}^\r {\wt
\ck}^\l}{(2{\sqrt \pi}^3)}\int_0^1 dx \,
e^{-\frac{i}{2}(2x-1)p\times p^{\prime}} \left[\frac{|{\wt
\ck}|}{2|\o_3|}\right]^{-1/2}K_{-1/2}(|{\wt \ck}||\o_3|),
    \ee
which in the rest frame gives the final answer in the form
    \be
\G_\m^{3NP} = \frac{e^2 \e_{\m\n\r}{p^\prime}^\n {\wt
\ck}^\r}{4\pi M} \left[\frac{1}{|{\wt \ck}|} - \frac{m}{2}
\right].
    \ee
In solving the momntum integral we have made use of the result
from Eq. (\ref{int4}). Similarly for the other two photon vertex
[Fig. (\ref{twop2})] we get,
    \be
\G_\m^{4NP}= - \frac{e^2 \e_{\m\n\r}p^\n {\wt \ck}^\r}{4\pi
M}\left[\frac{1}{|{\wt \ck}|}- \frac{m}{2}\right].
    \ee

\subsection{Two-photon and three-photon vertex}

\begin{figure}
\begin{center}
\includegraphics[width=2in,height=2in]{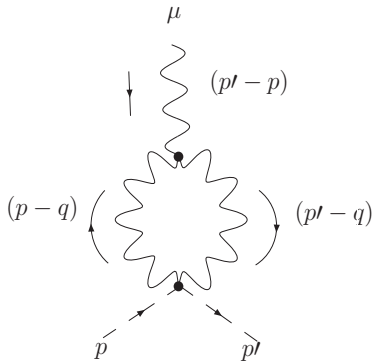}
\end{center}
\caption{The two and three photon vertex.} \label{twothreep}
\end{figure}
This last subsection deals with the two photon and three gluon
vertex. Similar to the contribution of Fig. (\ref{threep}) the
contribution from this diagram is purely due to NC nature of the
action. Calling the contribution from this diagram as $\G_\m^5$:
    \be
\G_\m^5 = \frac{4ie^2}{M}\intq {g_{\a\b} \e^{\a\n\l} \e_{\m\n\r}
\e^{\r\b\d} (p^{\prime}-q)_\d (p-q)_\l \over (p^{\prime}-q)^2
(p-q)^2} \cos\left[\frac{(p^{\prime}-q)\times (p-q)}{2} \right]
\sin\left[\frac{(p^{\prime}-p)\times (p^{\prime}-q)}{2}\right]
e^{\frac{i}{2} p \times p^{\prime}}.
    \ee
The standard manipulations give
    \be
\G_\m^5 = -\frac{2ie^2}{M} \intq {\e_{\m\n\l}\over
(p^{\prime}-q)^2 (p-q)^2} \left[\frac{\cp^\l \ck^\n}{2}+ \ck^\l
q^\n \right] \sin[p \times p^{\prime}-q\times \ck]
e^{\frac{i}{2}p\times p^{\prime}}.
    \ee
Proceeding as before we get
    \be
\G_\m^5 = \frac{2ie^2}{M} \int_0^1 dx \intq
\frac{\e_{\m\n\l}\ck^\l q^\n}{(q^2 - \o_5^2)^2} \sin[q \times \ck]
e^{\frac{i}{2} p \times p^{\prime}},
    \ee
where $\o_5^2=x(x-1) \ck^2$. Performing the momentum integration,
with $q={\bar q}+p+x \ck$, one gets
    \be
\G_\m^5 = -\frac{2e^2 \e_{\m\n\l}\ck^\l {\wt \ck}^\n}{(2
\sqrt{\pi})^3 M} \int_0^1 dx \left[\frac{|{\wt
\ck}|}{2|\o_5^2|}\right]^{-1/2} K_{-1/2}(|{\wt \ck}||\o_5^2|)
e^{\frac{i}{2} p \times p^{\prime}}.
    \ee
Upon simplification, the contribution from this vertex diagram
turns out to be
    \be
\G_\m^5 = \frac{e^2 \e_{\m\n\r}\ck^\n {\wt \ck}^\r}{4 \pi M |{\wt
\ck}|}.
    \ee

Combining the various vertices at first order in $\th$, one gets
        \be \label{result}
\G_\m= -{ie^2\e_{\m\n\r}\cp^\n \ck^\r \over 4\pi
M}\left[\frac{1}{m}- \frac{3|{\wt\ck}|}{4}\right]+ \frac{e^2
\e_{\m\n\r}\ck^\n {\wt \ck}^\r}{4\pi M}\left[\frac{2}{|{\wt \ck}|}
- \frac{m}{2} \right].
    \ee
The above vertex contributions, as can be noticed, is separates
into real and imaginary parts. The real part results due to the
appearance of the $\th$ dependent spin type term, unlike the other
term where only the magnitude of $\th$ appears. It can be seen
from the above expression that, the theta independent term of the
first piece arises due to the original vertex diagram which
yielded a finite value to the parity odd part of the vertex
function present in the commutative theory \cite{kogan1}. This
parity odd term can couple to the external magnetic field and
hence it was interpreted as the magnetic moment for the scalar
particles. The present term receives a finite NC correction due to
the appearance of non-planar integrals. This correction depends on
the value of the NC parameter and hence can also be interpreted as
a correction to the magnetic moment structure. The real piece of
the vertex function is interesting because, the parity odd spin
term couples not only to the external fields but it also couples
to the NC parameter. Similar structures also arises in the angular
momentum operator as has been shown recently \cite{mohrbach}.

\section{Conclusions}

In conclusion, we have evaluated the NC vertex diagrams at one
loop level, up to first order in $\th$, for the scalar particles.
The non-planar contributions brought in corrections to the spin
structure and also coupled the external field to the $\th$ tensor.
It has been shown recently that, the magnetic moment for scalar
matter fields in NC Maxwell-Chern-Simons can lead to the formation
of bound states on plane \cite{ghosh}. Hence, the implication of
these loop corrections needs careful investigation. $\th$
dependent contributions to angular momentum has been computed
recently and has been related to Berry's phase in momentum space.
It is worth reminding that, in the conventional CS theory the
Berry's phase has manifested in the context of Hall effect. In
this light, it will be exciting to study the NC contribution more
carefully.

When the present manuscript was under preparation, we came across
a preprint \cite{ajsilva}, where the authors have analyzed the
singularity structure of the vertex diagrams considered here.
\footnote{Private communication.}

\acknowledgments

T.S. acknowledges useful discussions with B. Chandrasekhar.

\appendix

\section{Important Integrals}

Below we give the solutions of the various loop integrals
encountered in the present work. For the sake of notational
simplicity we denote ${\wt\ck}^\m \equiv \th^{\m\n}{\ck}_\n$.
    \be \label{int2}
\frac{1}{ABC}= 2\int^{1}_{0}dx \int^{x}_0 dy \, [Ay + B(x-y) +
C(1-x)]^{-3}.
    \ee
    \be \label{int4}
\intq \frac{q_\m}{[q^2-\o^2]^2}\,e^{\pm iq\times \ck}=\mp
\frac{{\wt \ck}_\m}{(2\sqrt{\pi})^3}\left[\frac{|{\wt
\ck}|}{2|\o|}\right]^{-1/2}\,K_{-1/2}(|{\wt \ck}| |\o|).
    \ee
    \be \label{int8}
\intq \frac{q_\m q_\n e^{\pm iq\times \ck}}{[q^2-\o^2]^3}=
\frac{i}{(2\sqrt{\pi})^3} \left[\frac{{\wt \ck}_\m {\wt
\ck}_\n}{4} \left[\frac{|{\wt
\ck}|}{2|\o|}\right]^{-1/2}K_{-1/2}(|{\wt \ck}|
|\o|)-\frac{g_{\m\n}}{2} \left[\frac{|{\wt
\ck}|}{2|\o|}\right]^{1/2}K_{1/2}(|{\wt \ck}| |\o|) \right].
    \ee

\end{document}